\documentclass[preprint,aps,nofootinbib]{revtex4-1}
\pdfoutput=1

\usepackage{amsmath}  
\usepackage{amsbsy,amssymb,amsfonts}
\usepackage{graphicx}
\usepackage{color}
\usepackage{epsf}
\usepackage[utf8x]{inputenc}
\usepackage{hyperref}
\usepackage{multirow}
\def\ie{{\em{i.e.}},}

\def\bravert{\egroup\,\vrule\,\bgroup}

{\catcode`\|=\active
  \gdef\Twoint#1{\left(\mathcode`\|"8000\let|\bravert {#1}\right)}}
{\catcode`\|=\active
  \gdef\Braket#1{\left<\mathcode`\|"8000\let|\bravert {#1}\right>}}

\newcommand{\beq}{\begin{equation}}
\newcommand{\eeq}{\end{equation}}
\newcommand{\beqa}{\begin{eqnarray}}
\newcommand{\eeqa}{\end{eqnarray}}
\newcommand{\bea}{\begin{array}}
\newcommand{\eea}{\end{array}}

\newcommand{\bef}{\begin{figure}}
\newcommand{\ef}{\end{figure}}
\newcommand{\bc}{\begin{center}}
\newcommand{\ec}{\end{center}}
\newcommand{\bt}{\begin{table}}
\newcommand{\et}{\end{table}}
\newcommand{\btb}{\begin{tabular}}
\newcommand{\etb}{\end{tabular}}

\def\rvac{\left| \rule{0.3cm}{.0cm} \right>}

\begin{document}

\title{Electric Dipole Moments due to Nuclear Schiff Moment Interactions: \\
A Reassessment of the Atoms {$^{129}$Xe}, {$^{199}$Hg}, and the molecule {$^{205}$TlF}}

\vspace*{1cm}

\author{Micka\"el Hubert}
\email{mickael.hubert@epita.fr}
\affiliation{EPITA,
14 Rue Claire Pauilhac, 31000 Toulouse, France
}

\author{Timo Fleig}
\email{timo.fleig@irsamc.ups-tlse.fr}
\affiliation{Laboratoire de Chimie et Physique Quantiques,
             IRSAMC, Universit{\'e} Paul Sabatier Toulouse III,
             118 Route de Narbonne, 
             F-31062 Toulouse, France }

\vspace*{1cm}
\date{\today}

\vspace*{1cm}
\begin{abstract}
We present relativistic many-body calculations of atomic and molecular Schiff-moment interaction
constants including interelectron correlation effects using atomic Gaussian basis sets 
specifically optimized for the Schiff interaction. Our present best results employing a 
Gaussian nuclear density function are $\alpha_{\text{SM}} =
(0.364 \pm 0.025) \times 10^{-17} \frac{e \text{cm}}{e \text{fm}^3}$ for atomic {$^{129}$Xe},   
$\alpha_{\text{SM}} =
(-2.40 \pm 0.24) \times 10^{-17} \frac{e \text{cm}}{e \text{fm}^3}$ for atomic {$^{199}$Hg},   
and $W_{\text{SM}} = (39967\pm 3600)$ a.u. for the thallium nucleus in the molecule {$^{205}$TlF}.
We discuss agreements and discrepancies between our present results and those from earlier
calculations on the atoms {$^{129}$Xe} and {$^{199}$Hg}.
Using the most recent measurements of ${\cal{P,T}}$-odd electric dipole moments and the present
interaction constants reliable upper bounds on the Schiff moments of the 
{$^{199}$Hg} and {$^{205}$Tl} nuclei are determined in the context of a single-source assumption.
\end{abstract}

\maketitle
\section{Introduction}
\label{SEC:INTRO}

The Standard Model (SM) of elementary particle physics \cite{Glashow_1961EW,Weinberg_1967Leptons,Salam:1968rm}
is an extremely well-tested theory of fundamental particles and their interactions.
However, it leaves a number
of firmly established observations about our universe unexplained, like its matter and energy content
\cite{Hinshaw:2012aka}. Specifically, the SM does not allow to accomodate baryon asymmetry of the
universe (BAU), \ie\ the significant surplus of denominational matter over antimatter in the universe
\cite{Dine_Kusenko_MatAntimat2004}.

A necessary condition for explaining the BAU is the violation of the combined discrete symmetries
charge conjugation and parity (${\cal{CP}}$),
favoring the production of a matter over an antimatter particle \cite{Sakharov_JETP1967}.
${\cal{CP}}$-violation (CPV) has been observed in the decay of certain mesons \cite{K-meson,B-meson,Abe:2001xe},
and a partial theory of ${\cal{CP}}$-violation has become an integral part of the SM \cite{Kobayashi}.
However, it is generally agreed that this manifestation of ${\cal{CP}}$-violation is insufficient for
explaining the BAU. New sources of ${\cal{CP}}$-violation are hence required, which in particular generally
also yield flavor-diagonal ${\cal{CP}}$-violation \cite{FlavorPhysLepDipMom_EPJC2008}.
Under the assumption that ${\cal{CPT}}$ invariance (${\cal{T}}$ denoting time reversal)
of fundamental physical laws holds \cite{pauli_lorentz_CPT}, CPV implies the violation of
${\cal{T}}$ symmetry. The detection of an atomic or molecular electric dipole moment (EDM),
the Hamiltonian of which is ${\cal{P,T}}$-odd, would in turn indicate CPV. Thus, EDMs represent a
nearly background-free probe of beyond-SM CPV \cite{ramsey-musolf_review1_2013}.

The origins of nuclear, atomic and molecular EDMs in terms of fundamental CPV phases may be diverse
\cite{EDMsNP_PospelovRitz2005,chupp_rmp_2019,Shindler_EDMreview_2021}. In electronically
closed-shell systems like atomic mercury (Hg) \cite{swallows_Hg_PRA2013} or molecular thallium fluoride
(TlF) \cite{CENTREX_2021,Flambaum-Dzuba_TranPRA2020} the
situation is relatively complicated, and several hadronic mechanisms may contribute to leading
order. The nuclear Schiff moment and the nucleon-electron tensor-pseudotensor interaction are
the leading CPV sources at the hadronic and nuclear energy scale in such systems.

The nuclear Schiff potential is a low-order ${\cal{P,T}}$-odd term in the expansion of the
nuclear charge distribution \cite{khriplovich_lamoreaux,spevak_auerbach_flambaum1997} which polarizes
the electron cloud in an atomic system, giving rise to atomic/molecular EDMs. Thus, atomic-scale
measurements search for or constrain the nuclear Schiff moment and -- in turn -- the underlying
CPV sources. The chosen systems and states have electronically closed shells which strongly
suppresses leptonic CPV sources such as the electron EDM, $d_e$, and some semi-leptonic CPV sources 
such as the nucleon-electron scalar-pseudoscalar coupling, $C_S$ 
\cite{Barr_eN-EDM_Atoms_1992,Fleig:2018bsf}.
The currently most sensitive measurement \cite{Heckel_Hg_PRL2016} is
$|d_{\rm{Hg}}| < 7.4 \times 10^{-30} e$cm with spin-polarized mercury atoms, {$^{199}$Hg},
leading to an upper bound on the Schiff moment of $|S_{\rm{Hg}}| < 3.1 \times 10^{-13} e$fm$^3$
($95$\% c.l.). In the calculation of this upper bound an average over interaction constants
from different theory groups has been used.
The upper bound translates \cite{swallows_Hg_PRA2013,Flambaum-Dzuba_PRA2020} into constraints on
more fundamental parameters, the CPV pion-nucleon couplings $g_{\pi}^{(0)}$, $g_{\pi}^{(1)}$,
$g_{\pi}^{(2)}$, the nucleon EDMs $d_p$ and $d_n$, the quantum chromodynamics $\Theta$
parameter and chromo-EDMs.

In the present paper we pursue several goals:
\begin{enumerate}
%
 \item The Schiff moment interaction in general is quite strongly dependent on the quality of the 
	atomic basis set for the target nucleus in electronic-structure calculations, as has also been 
	substantiated earlier \cite{Quiney_PTodd_PRA1997,Petrov_TlF_PRL2002}. Our calculations confirm this 
	finding for all systems we have studied so far. In the present paper present a systematic approach to extending standard atomic basis sets with the aim
       of a reliable and economic description of the Schiff moment interaction in both atoms and molecules.
       In the following section we define our method for calculation of atomic and molecular Schiff-moment
	interactions and give a detailed description of the strategy for optimizing required Gaussian basis
	sets.
 \item In section III we discuss applications to {$^{129}$Xe}, {$^{199}$Hg} and $^{205}$TlF using a large
	 optimized basis set for each respective system and carefully taking into account interelectron
	correlation effects shellwise and at various excitation ranks.
	 Based on our findings we address current disagreements among previously published results 
	for atomic Schiff moment interactions in {$^{129}$Xe} and {$^{199}$Hg}.
 \item  We conclude on our study in section IV and use the most recent experimental EDM measurements 
	 \cite{Heckel_Hg_PRL2016,ChoSangsterHinds_TlF_PRA1991} and our calculated interaction 
	constants for these systems to derive constraints on the respective nuclear Schiff moments.
\end{enumerate}

\section{Theory and Methods}
\label{SEC:THEORY}
\subsection{Theory}

\subsubsection{The Atomic Schiff Moment Interaction Hamiltonian}

The atomic Schiff-moment interaction for a single electron in the field of a point nucleus has been given
as \cite{khriplovich_lamoreaux}, Eq. (8.75),
%
\begin{equation}
	{\hat{H}}_{\text{SM}} = -e\, {\bf{S}} \cdot {\boldsymbol{\nabla}}_r \delta({\bf{r}})
	\label{EQ:SCHIFF_HAM_DELTA}
\end{equation}
where the vector coefficient ${\bf{S}} := S\, \frac{{\bf{I}}}{I}$ is the Schiff moment \cite{flambaum_ginges2002}
of the nucleus with ${\bf{I}}$ denoting nuclear spin and $S$ the scalar Schiff-moment constant
\cite{Schiff_PRA_2002}.

Recent developments use a more realistic finite nuclear charge density,
and the Hamiltonian for the interaction 
of an electron with the Schiff potential \cite{Schiff_PRA_2002} has been represented as
\begin{equation}
	{\hat{H}}_{\text{SM}} = -e \varphi_{\text{SM}}({\bf{r}}) = -3e\, \frac{{\bf{S}}\cdot{\hat{\bf{r}}}}{B} \rho({\bf{r}})
	\label{EQ:SCHIFF_HAM}
\end{equation}
where $B = \int\limits_0^{\infty}\, \rho({\bf{r}}) r^4 dr$, and $\rho({\bf{r}})$ the nuclear charge density 
at position ${\bf{r}}$.

\subsubsection{Expectation value approach to Schiff moment interaction}

Our principal strategy is to determine the $E$-field dependent ${\cal{P,T}}$-odd energy shift 
$\Delta \varepsilon$ as a function of an atomic interaction constant. The full details of the 
general approach are found in Ref. \cite{Fleig_Skripnikov2020}. Here, we present the specific 
approach for the calculation of Schiff-moment interaction constants.

\noindent
\paragraph{Atoms}

For atomic calculations we include an external homogeneous electric field $E_{\text{ext}}$ 
along the $z$ axis.
In an atom with $n$ electrons the associated ${\cal{P,T}}$-odd energy shift can then be 
expressed as an expectation value over the one-electron Hamiltonian in Eq. (\ref{EQ:SCHIFF_HAM}) (here in a.u.)
\begin{equation}
 \Delta \varepsilon_{\text{SM}} = -S_z \frac{3}{B}\, \left< \sum\limits_{j=1}^n\, {\hat{z}_j}\, \rho({\bf{r}}_j) \right>_{\psi(E_{\text{ext}})} 
 \label{EQ:SCHIFF_EXP}
\end{equation}
where $\psi(E_{\text{ext}})$ is the electronic wavefunction of the field-dependent state. 
We first solve a zeroth-order problem
\begin{equation}
 \hat{H}({{E}_{\text{ext}}}) \big| \psi(E_{\text{ext}}) \big>
	= \varepsilon({{E}_{\text{ext}}}) \big| \psi(E_{\text{ext}}) \big>
\end{equation}
with $\varepsilon$ the field-dependent energy eigenvalue and $\hat{H}({{E}_{\text{ext}}})$ the
Dirac-Coulomb Hamiltonian including the interaction term with the external field:
\begin{eqnarray}
 \nonumber
 \hat{H}({{E}_{\text{ext}}}) &:=& \hat{H}^{\text{Dirac-Coulomb}} + \hat{H}^{\text{Int-Dipole}} \\
  &=& \sum\limits^n_j\, \left[ c\, \boldsymbol{\alpha}_j \cdot {\bf{p}}_j + \beta_j c^2 
- 
  \frac{Z}{r_{jK}}{1\!\!1}_4 \right]
+ \sum\limits^n_{k>j}\,
\frac{1}{r_{jk}}{1\!\!1}_4
  + \sum\limits_j\, {\bf{r}}_j \cdot {\bf{E}_{\text{ext}}}\, {1\!\!1}_4
 \label{EQ:HAMILTONIAN}
\end{eqnarray}
${\bf{E}_{\text{ext}}}$ is weak (see below for details) and homogeneous,
the indices $j,k$ run over $n$ electrons, $Z$ is the proton 
number with the nucleus $K$ placed at the origin, and $\boldsymbol{\alpha}, \beta$ are standard 
Dirac matrices.
$E_{\text{ext}}$ is not treated as a perturbation but included {\it{a priori}} in the variational 
optimization of the wavefunction, $\psi(E_{\text{ext}})$.

Technically, $\psi(E_{\text{ext}})$ is a configuration interaction (CI) vector \cite{knecht_luciparII} built from 
Slater determinants over field-dependent 4-spinors.
In the atomic case the wavefunction is expanded as follows:
\begin{equation}
 \psi(E_{\text{ext}})\, \widehat{=}\, \left| M_J \right> = \sum\limits_{I=1}^{\rm{dim}{\cal{F}}^t(M,n)}\,
                                       c_{(M_J),I}\, ({\cal{S}}{\overline{\cal{T}}})_{(M_J),I} \rvac
        \label{EQ:AT_WF}
\end{equation}
where $\rvac$ is the true vacuum state,
${\cal{F}}^t(M,n)$ is the symmetry-restricted sector of Fock space ($M_J$ subspace) with $n$ electrons in 
$M$ four-spinors,
${\cal{S}} = a^{\dagger}_i a^{\dagger}_j a^{\dagger}_k \ldots$ is a string of spinor creation operators,
${\overline{\cal{T}}} = a^{\dagger}_{\overline l} a^{\dagger}_{\overline m} a^{\dagger}_{\overline n} \ldots$
is a string of creation operators of time-reversal transformed spinors. The determinant expansion coefficients
$c_{(M_J),I}$ are generally obtained as described in refs. \cite{fleig_gasci,fleig_gasci2}.

The electric dipole moment of the atomic system in terms of the Schiff moment is
\begin{equation}
	d_a = \alpha_{\text{SM}}\, S_z
 \label{EQ:ATOMIC_EDM_SCHIFF}
\end{equation}

where using Eqs. (\ref{EQ:SCHIFF_EXP}) and (\ref{EQ:ATOMIC_EDM_SCHIFF}) we define the Schiff 
moment interaction constant
\begin{equation}
	\alpha_{\text{SM}} := \frac{\Delta \varepsilon_{\text{SM}}}{S_z\, E_{\text{ext}}}
			    = \frac{-\frac{3}{B}\, \bigg< \sum\limits_{j=1}^n\, {\hat{z}_j}\, \rho({\bf{r}}_j) \bigg>_{\psi(E_{\text{ext}})}}{E_{\text{ext}}}.
 \label{EQ:ATOMIC_SCHIFF_CONSTANT}
\end{equation}

In the linear r\'egime $W_a(\psi(E_{\text{ext}})) 
:= -\frac{3}{B}\, \big< {\hat{z}}\, \rho({\bf{r}}) \big>_{\psi(E_{\text{ext}})} = C\, E_{\text{ext}}$ where 
$C$ is a constant with respect to $E_{\text{ext}}$. 
From this it follows that in
the linear r\'egime $\alpha_{\text{SM}} =C$ which is independent of $E_{\text{ext}}$. Through numerical
analysis we determine quasi linearity of $W_a$ for $E_{\text{ext}} = 0.0003$ a.u. in the case of Xe and
for $E_{\text{ext}} = 0.00024$ a.u. in the case of Hg. $E_{\text{ext}}$ is oriented along the $z$ axis in
the atomic case. 

\noindent
\paragraph{Molecules}
In the molecular case the unperturbed wavefunction $\psi$ is not a ${\cal{P}}$ eigenstate, thus no external 
electric field needs to be applied. The general strategy is similar to the one for atoms but with some 
modifications. The energy shift is written as
\begin{equation}
 \Delta \varepsilon_{\text{SM}} = -S_z \frac{3}{B}\, \bigg< \sum\limits_{j=1}^n\, {\hat{z}_j}\, \rho({\bf{r}}_j) \bigg>_{\psi}
 \label{EQ:SCHIFF_EXP_MOL}
\end{equation}
and the wavefunction is obtained from the zeroth-order problem
\begin{equation}
 \hat{H} \big| \psi \big>
        = \varepsilon \big| \psi \big>
\end{equation}
with
\begin{eqnarray}
 \nonumber
 \hat{H} &:=& \hat{H}^{\text{Dirac-Coulomb}}  \\
  &=& \sum\limits^n_j\, \left[ c\, \boldsymbol{\alpha}_j \cdot {\bf{p}}_j + \beta_j c^2
-
  \sum\limits^2_K\, \frac{Z_K}{r_{jK}}{1\!\!1}_4 \right]
+ \sum\limits^n_{k>j}\,
\frac{1}{r_{jk}}{1\!\!1}_4 + V_{KL}
 \label{EQ:HAMILTONIAN_MOL}
\end{eqnarray}
for a diatomic molecule
where $K$ runs over nuclei and $V_{KL}$ is the classical electrostatic potential energy for
the two Born-Oppenheimer-fixed nuclei. The CI expansion of the electronic wavefunction reads
\begin{equation}
 \psi\, \widehat{=}\, \left| \Omega \right> = \sum\limits_{I=1}^{\rm{dim}{\cal{F}}^t(M,n)}\,
                                       c_{(\Omega),I}\, ({\cal{S}}{\overline{\cal{T}}})_{(\Omega),I} \rvac
        \label{EQ:MOL_WF}
\end{equation}
where $\Omega$ is the total angular momentum projection. The Schiff-moment interaction constant for a target
nucleus $A$ of a molecule is then written as
\begin{equation}
	W_{\text{SM}}(A) := \frac{\Delta \varepsilon_{\text{SM}}(A)}{S_z(A)}
		    = -\frac{3}{B}\, \bigg< \sum\limits_{j=1}^n\, {\hat{z}_j}\, \rho_A({\bf{r}}_j) \bigg>_{\psi}.
 \label{EQ:MOLECULAR_SCHIFF_CONSTANT}
\end{equation}
In practical applications the target nucleus is placed at the origin of the reference frame.

\subsection{Methods}

\subsubsection{Nuclear charge density}

To describe the charge density, $\rho({\bf r})$, at position ${\bf r}$ for a nucleus with $Z$ protons 
we in the present work use a Gaussian model \cite{Visscher_Dyall_nuclcha} with
\begin{equation}
	\rho({\bf r}) = Z \left(\frac\zeta\pi\right)^{\frac 32} \, e^{-\zeta {\bf r}^2}
 \label{EQ:Gaurho}
\end{equation}
where the exponent $\zeta$ is taken from Ref. \cite{Visscher_Dyall_nuclcha}. 
This density is used both for the calculation of the electronic wavefunction as well as for the evaluation
of the interaction constants in Eqs. \ref{EQ:ATOMIC_SCHIFF_CONSTANT} and \ref{EQ:MOLECULAR_SCHIFF_CONSTANT}.

\subsubsection{Matrix Elements}
We here present the main ideas of the Schiff-moment interaction operator implementation. 
In a Gaussian nuclear model the prefactor $B$ introduced in Eq. (\ref{EQ:SCHIFF_HAM}) can be written 
in terms of the parameter $\zeta$ from Eq. (\ref{EQ:Gaurho}). 
Integration by parts leads to
\begin{equation}
	B = \frac 3{8\pi\zeta}.
 \label{EQ:BINT}
\end{equation}
The electronic spinors constituting the wavefunction $\psi$ in Eqs. 
(\ref{EQ:ATOMIC_SCHIFF_CONSTANT}) and (\ref{EQ:MOLECULAR_SCHIFF_CONSTANT}) 
are expanded as a linear combination of primitive Gaussians in the DIRAC code. It is, therefore,
convenient to contract with the exponential $e^{-\zeta {\bf r}^2}$ from Eq. (\ref{EQ:Gaurho})
by adding $\zeta$ to the primitive Gaussian exponents.
Then, the matrix elements to be evaluated in Eqs. (\ref{EQ:ATOMIC_SCHIFF_CONSTANT}) 
and (\ref{EQ:MOLECULAR_SCHIFF_CONSTANT}) can be written as
\begin{equation}
	\left<\psi\left| e^{-\zeta {\bf r}^2} \hat z \right|\psi\right> =  \left<\psi_\zeta \left| \hat z \right| \psi \right> 
 \label{EQ:ZRHO}
\end{equation}
where $\psi_\zeta = \psi e^{-\zeta {\bf r}^2}$. Finally, the r.h.s. of Eq. (\ref{EQ:ZRHO}) is evaluated
as a dipole length integral in the DIRAC package.

\subsubsection{Atomic basis sets}

The nuclear Schiff moment gives rise to an asymmetric charge distribution on the nuclear surface and a
related constant electric field inside the nucleus \cite{flambaum_ginges2002} that is oriented along the 
nuclear spin, ${\bf{I}}$. The atomic Schiff moment interaction will, therefore, predominantly affect atomic
electronic wavefunctions that penetrate the atomic nucleus. Eq. (\ref{EQ:SCHIFF_HAM}) shows that the
Schiff moment interaction is ${\cal{P}}$- and ${\cal{T}}$ violating, thus leading to a mixing of states
with opposite parity, and therefore predominantly to mixing of electronic $s$ and $p$ waves. 

Our first goal was to investigate the performance of the standard systematic N-tuple-zeta series of Gaussian
basis sets in calculating the Schiff-moment interaction constant $\alpha_\text{SM}$ in Dirac-Coulomb
Hartree-Fock (DCHF) approximation. The results in Table \ref{TAB_Xe:final-basis} demonstrate that
$N=4$ is a minimal requirement for quantitatively reliable results. Still, compared to the literature
results that agree well in mean-field approximation, a standard QZ basis set yields a too small result.
Gaussian basis sets of quintuple-zeta (and higher) quality are currently not available for heavy atoms
and are time consuming to develop \cite{dyall_s-basis}.
Given the physical nature of the Schiff-moment interaction it is, therefore, as an alternative 
strongly suggested to increase the $s$ 
and $p$ subspaces of the most extensive standard Gaussian basis set in order to obtain an accurate description of the relevant 
matrix elements, written generically as 
\begin{equation}
	\label{EQ:ME_GEN_SCHIFF}
	\left< s \left| -\frac{3}{B}\, {\hat{z}}\, \rho({\bf{r}}) \right| p \right>.
\end{equation}

The strategy of our present basis-set optimization starts from Dyall's relativistic Gaussian basis set, QZ, extended 
with diffuse and correlating functions \cite{4p-basis-dyall-2}. This set is then further augmented taking
two criteria into account:

\begin{enumerate}
 \item Densification in the $s$ and $p$ spaces.  \\
       Following the so-called ``even-tempered prescription'' (see Ref. \cite{dyfae_corr_schwerdt}) we 
       insert a Gaussian function between each adjacent pair of original ones according to
       \begin{equation}
	\zeta_{n} = \sqrt{\zeta_{n-1} \, \zeta_{n+1}} 
        \label{even-tempered}
       \end{equation}
       where $\zeta_i$ is the exponent in $e^{-\zeta_{i} {\bf r}^2}$ of the $i^{\rm th}$ Gaussian function
       and $\zeta_{n-1} (\zeta_{n+1})$ is the next larger (smaller) exponent. This procedure could in principle
       be repeated several times but the rapid increase in dimension of the $s$ and $p$ spaces leads to a
       steep increase in computational cost.

 \item Addition of $sp$ pair(s) of even-tempered dense and diffuse functions.

In order to obtain a more extended basis set in a balanced way we add a pair of Gaussian functions -- one diffuse
and one dense -- to the respective densified basis set. The new compact exponent $\zeta_{n+1}$ is 
obtained according to
\begin{equation}
	\zeta_{n+1} = \frac{\zeta_{n}^2}{\zeta_{n-1}}
\label{even-tempered2}
\end{equation}
where $\zeta_{n}$, $\zeta_{n-1}$ are the two most compact coefficients 
in the $sp$-densified basis-set defined in the latter point 1. 

The new diffuse exponent is obtained accordingly:
\begin{equation}
	\zeta_{n-1} = \frac{\zeta_{n}^2}{\zeta_{n+1}}
\label{even-tempered3}
\end{equation}
where $\zeta_{n-1}$ is the new more diffuse coefficient and $\zeta_{n}$, $\zeta_{n+1}$ are the 
two most diffuse coefficients in the $sp$-densified 
basis set defined in point 1.
\end{enumerate}

\begin{figure}[h]
	\caption{\label{FIG_Xe:hf(nsp)} $\varepsilon_{\text{DCHF}}$ variation from sp-densified QZ for Xe. }
 \includegraphics[angle=0.,width=15.0cm]{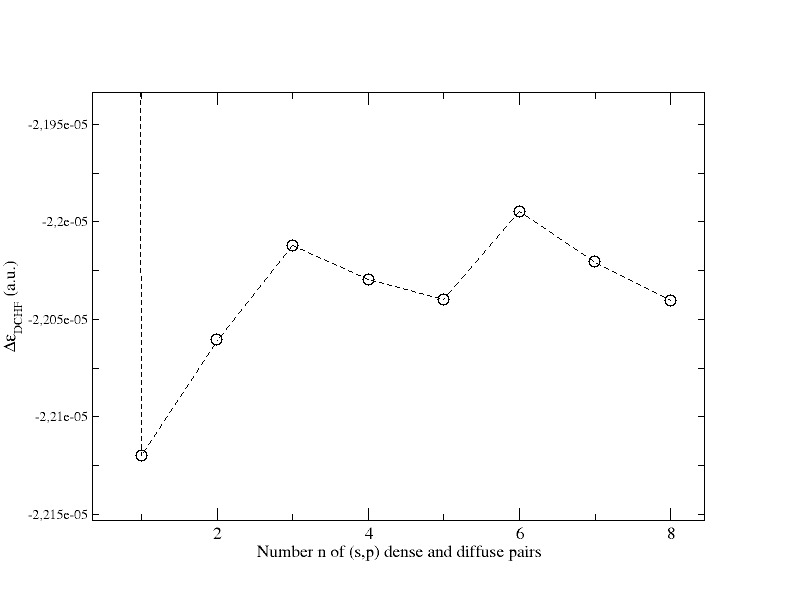}
\end{figure}

As shown in Table 
\ref{TAB_Xe:final-basis} for Xe neither is the total Dirac-Coulomb Hartree-Fock energy 
(${\varepsilon_\text{DCHF}}$) improved (lowered) nor is the Schiff moment interaction constant $\alpha_{\text{SM}}$
changed substantially by a second densification if the respective densification is accompanied by 
the addition of a sufficient number of pairs of even-tempered compact and diffuse functions 
(respectively, +$1sp$ and +$3sp$ for simple and double densification, see Table \ref{TAB_Xe:nsp-densified}). 
For this reason, we densify the original basis set only once. 
From Tables \ref{TAB_Xe:final-basis}, \ref{TAB_Hg:final-basis}, \ref{TAB_Xe:nsp-densified} and 
Fig. \ref{FIG_Xe:hf(nsp)}, 
respectively, we conclude that our accurate and most economic basis set 
to describe $\alpha_\text{SM}$ for Xe and Hg is the single $sp$-densified Dyall's cvQZ + $1sp$. 

Combining the two aforementioned criteria, we propose a systematic method to optimize Gaussian basis 
sets suited to address the Schiff moment interaction constant $\alpha_\text{SM}$.
The starting point is an even-tempered $(s,p)$-spaces densification. Then, compact and diffuse 
$(s,p)$ pairs are added until a minimal DCHF energy converged at a level of 
$\approx 10^{-6}$ a.u. in reached, as is shown in Fig. \ref{FIG_Xe:hf(nsp)}. 
Under these circumstances $\alpha_\text{SM}$ shall be considered converged at DCHF level.

\subsubsection{Molecular basis sets}

Obtaining a suitable basis set for a target atom in a diatomic molecule can be achieved by following steps 
1 and 2 described in the latter atomic case for the atom with the target nucleus A 
(see Eq. (\ref{EQ:MOLECULAR_SCHIFF_CONSTANT})). 
However the basis-set optimization must be done by calculating ${\varepsilon_\text{DCHF}}$ and 
$W_{\text{SM}}(A)$ in the molecular field.
The internuclear distance $R$ is kept fixed during the whole process. It is obtained from experimental data in the present
case.
For TlF, we conclude from Table \ref{TAB_TlF:final-basis} 
that our accurate and most economic basis set to describe $W_{\text{SM}}(A)$ is the $sp$-densified Dyall's cvQZ + $1sp$.

\section{Results for Schiff moment interaction}
\label{SEC:APPL}
\subsection{Technical Details}
All present calculations have been carried out using a locally-modified version of the DIRAC 
program package \cite{DIRAC_JCP}. The chosen symmetry group is the double point group $C_{2v}^*$.
Small components of the Dirac spinors are generated through the restricted-kinetic-balance
\cite{kin_bal}
prescription and all small-component integrals are explicitly taken into account.

\subsection{{$^{129}$Xe}}

\begin{table}[h]
 \caption{\label{TAB_Xe:final-basis} Atomic Schiff moment interaction constant for Xe calculated at Hartree-Fock level
	including core contribution with Gaussian nuclear density \cite{Visscher_Dyall_nuclcha} 
	for studying electronic-basis-set convergence; $E_{\text{ext}}$ is set to $0.0003$ a.u.
        Augmented basis sets are built from Dyall's QZ set including diffuse and correlating functions.\\ 
           }

 \vspace*{0.3cm}
 \hspace*{-1.3cm}
\begin{tabular}{l|cc}
	Model & $\alpha_{\text{SM}}~ \left[ 10^{-17} \frac{e \text{cm}}{e \text{fm}^3} \right]$ & $\varepsilon_{\text{DCHF}}$ [a.u.]   \\ \hline
	DZ-21s15p                                        &  $-1.220$   & $-7446.876435682$  \\
	TZ-29s22p                                        &  $-0.379$   & $-7446.895053545$  \\
	QZ-34s28p                                              &  $ 0.318$   & $-7446.895409376$  \\
	sp-densified QZ-67s55p                                 &  $ 0.314$   & $-7446.895379750$  \\\hline
	sp-densified+1sp QZ-69s57p                             &  $ 0.373$   & $-7446.895401869$  \\ \hline
	sp-densified+2sp QZ-71s59p                             &  $ 0.375$   & $-7446.895401810$  \\ 
	sp-densified+3sp QZ-73s61p                             &  $ 0.375$   & $-7446.895401761$ \\ 
	sp-densified+4sp QZ-75s63p                             &  $ 0.375$   & $-7446.895401779$ \\ 
	Double sp-densified QZ-133s109p                        &  $ 0.362$   &$-7446.895392349$  \\ 
	Double sp-densified+3sp QZ-139s115p                    &  $ 0.369$   &$-7446.895401848$  \\ \hline 
	Dzuba {\it{et al.}}\cite{Schiff_PRA_2002} (RPA, 2002)  &                    \multicolumn{1}{c}{$0.38$} & -\\
	Ramachandran {\it{et al.}}\cite{Ramachandran-Latha_PRA2014} (CPHF, 2014) & \multicolumn{1}{c}{$0.374$} &-\\
	Sakurai {\it{et al.}}\cite{Sakurai:2019vjs} (CPDF, 2019)  &               \multicolumn{1}{c}{$0.38$}& -

\end{tabular}
\end{table}
Previous calculations have been carried out in random-phase approximation (RPA) \cite{Schiff_PRA_2002}
and within the coupled-perturbed Dirac-Hartree-Fock (CPHF) framework \cite{Ramachandran-Latha_PRA2014}
yielding very similar results. 
In recent work using the relativistic normalized coupled cluster method 
in singles and doubles approximation (RNCCSD) \cite{Sakurai:2019vjs} interelectron correlation effects 
have been taken into account and a contribution of $\approx -15$\% to 
$\alpha_{\text{SM}}$ is reported which is unexpectedly large for Xe. Our general-excitation-rank CI 
method can test this claim.

\begin{table}[h]
 \caption{\label{TAB:SCH_XE} Atomic Schiff moment interaction constant for Xe including electron correlation
	effects and the core contribution using the Dyall-cvQZ-69s57p basis set and a Gaussian nuclear density 
	\cite{Visscher_Dyall_nuclcha}; $E_{\text{ext}}$ is set to $0.0003$ a.u.
           }

 \vspace*{0.3cm}
 \hspace*{-1.3cm}
\begin{tabular}{l|c}
CI Model/virtual cutoff & $\alpha_{\text{SM}}~ \left[ 10^{-17} \frac{e \text{cm}}{e \text{fm}^3} \right]$ \\ \hline
	DCHF                         &  $0.373$   \\
	SD8/5au                      &  $0.348$   \\
	SD8/10au                     &  $0.353$   \\
	SD8/20au                     &  $0.352$   \\
	SD8/50au                     &  $0.352$   \\ 
	SDT8/10au                    &  $0.351$   \\ 
	SDTQ\_0.2au\_SDT8/10au       &  $0.355$   \\ \hline
	S10\_SD18/10au               &  $0.355$   \\
	S10\_SD18/20au               &  $0.359$   \\
	S10\_SD18/50au               &  $0.359$   \\ 
	SD10\_SD18/20au              &  $0.362$   \\ \hline
	S8\_SD16/20au                &  $0.352$   \\ 
	S8\_SD16/50au                &  $0.353$   \\ 
	SD8\_SD16/20au               &  $0.352$   \\ \hline
	S8\_SD8\_SD24/20au           &  $0.352$   \\ \hline
	{\bf{Final present}}         &  {\bf{0.364 $\pm$ 0.025}}  \\ \hline
	Dzuba {\it{et al.}}\cite{Schiff_PRA_2002} (RPA, 2002)                    & $0.38$ \\
	Dzuba {\it{et al.}}\cite{dzuba_flambaum_PRA2009} (RPA, 2009)             & $0.38$ \\
	Ramachandran {\it{et al.}}\cite{Ramachandran-Latha_PRA2014} (CPHF, 2014) & $0.374$\\
	Sakurai {\it{et al.}}\cite{Sakurai:2019vjs} (CPDF, 2019)                 & $0.38$ \\ 
	Sakurai {\it{et al.}}\cite{Sakurai:2019vjs} (RNCCSD, 2019)               & $0.32$  
\end{tabular}
\end{table}
In Table \ref{TAB:SCH_XE} the results from a series of systematic calculations including electron 
correlation effects from various atomic shells and at various maximum excitation ranks are compiled.
As has also been found earlier \cite{dzuba_flambaum_PRA2009} even the leading 
correlation effects from the valence shells ($5s,5p$), described by Double excitations, are weak in
the ground state of atomic Xe. In our model SD8 they
decrease $\alpha_{\text{SM}}$ by only around $6$\%. The model SDT8 introduces all Triple excitations
and the model SDTQ\_0.2au\_SDT8 in addition a subset of Quadruple excitations (where the spinor
space for these Quadruples has been truncated at $0.2$ a.u.) to the model SD8. 
Including these higher excitation ranks affects the Schiff interaction constant by less than $1$\%.
Excitations out of the $4d$ shell lead to an increase by about $3$\%. Here we have not tested higher
excitation ranks than Doubles since the difference from adding Single and Double excitations
out of the $4d$ shell is already small (around $1$\%). Finally, the correlation contributions from core 
shells $3s,3p,4s,4p$ are seen to be smaller than $1$\%.

The final present value is thus calculated as follows. As base value we take the result where the
greatest number of electrons has been included in the correlation expansion, from the model 
S8\_SD8\_SD24/20au. To this we add corrections due to higher excitation ranks in the valence shells
and correlations among and with the $4d$ electrons, according to
\begin{eqnarray*}
\alpha_{\text{SM}}{\text{(Final)}} &=& \alpha_{\text{SM}}({\text{S8\_SD8\_SD24/20au)}} \\
 &+& \alpha_{\text{SM}}{\text{(SDTQ\_0.2au\_SDT8/10au)}}  - \alpha_{\text{SM}}{\text{(SD8/10au)}} \\
 &+& \alpha_{\text{SM}}{\text{(SD10\_SD18/20au)}}  - \alpha_{\text{SM}}{\text{(SD8/20au)}}
\end{eqnarray*}
To this final value we assign an uncertainty of $7$\% by adding individual uncertainties due to the
nuclear-density model ($3$\%), atomic basis set ($2$\%), higher excitation ranks ($1$\%), and the
Breit interaction ($1$\%) that is not present in our Hamiltonian, Eq. (\ref{EQ:HAMILTONIAN}).
Our final result including the error estimate is not in accord with the coupled cluster result from Ref. 
\cite{Sakurai:2019vjs}. However, in the latter work a final uncertainty estimate is not given.
According to V. Dzuba \cite{dzuba_priv} correlation contributions beyond the Random Phase
Approximation (RPA) in the case of the Xe atom are not greater than $\approx 3$\%. Our final
result indeed shows small total correlation effects and is in agreement with the results from Refs. 
\cite{Schiff_PRA_2002} and \cite{dzuba_flambaum_PRA2009}.


\subsection{{$^{199}$Hg}}

Table \ref{TAB:SCH_HG} shows correlated results for the mercury atom. Including only Single 
and Double excitations for the $12$ outermost electrons (shells $5d$ and $6s$) yields a correlation
contribution of roughly $10$\% on top of the DCHF value. This is a significantly greater contribution 
than the
corresponding one in atomic xenon. However, the model SD12 is still not sufficient. Adding full
Triple excitations to the wavefunction expansion results in a further $6.5$\% decrease of 
$\alpha_{\text{SM}}$ on the absolute. Comparing the model SDT12 with the more approximate expansion
SD10\_SDT12 shows that the effect of $3$ holes in the $5d$ spinor space is rather unimportant (only
$0.2$\% of the DCHF value) and that it is the combined higher excitations out of the $5d$ and $6s$
shells that have to be accounted for. We accomplish this through the model SD10\_SDTQ12 where the
excitation rank for a maximum of $2$ holes in the $5d$ spinors is maximal. This model yields another
$2.4$\% decrease, on the absolute, at a cutoff of $5$ a.u. for the virtual spinors.

Additional effects on $\alpha_{\text{SM}}$ from excitations out of the atomic core spinors are
accounted for using a virtuals cutoff of $20$ a.u.  We find that one- and two-hole configurations
in the $5p$ shell (model SD6\_SD18) contribute a mere $0.5$\%. On the other hand, one- and two-hole 
configurations in the $4f$ shell (model S20\_SD32) contribute about $3$\%. It is noteworthy that
these two corrections are opposed to each other: Excitations out of $p$ shells decrease electron
density in $p$-shell configurations that contribute directly to the generic Schiff 
moment interaction matrix element in Eq. (\ref{EQ:ME_GEN_SCHIFF}). This leads to an absolute decrease
of $\alpha_{\text{SM}}$. On the other hand, excitations out of the $f$ shell reduce the screening
of nuclear charge on electrons in directly contributing shells, and so lead to an absolute increase
of $\alpha_{\text{SM}}$. The same effect can be observed when the $5d$ shell is opened for excitations
(model SD12 vs. SD2).

We, therefore, obtain our final value from a base value with the largest number of correlated electrons
(SD34) improved by a correction for higher combined excitations (Triples and Quadruples) from the valence 
shells, according to
\begin{eqnarray*}
\alpha_{\text{SM}}{\text{(Final)}} &=& \alpha_{\text{SM}}({\text{SD34/20au)}} \\
 &+& \alpha_{\text{SM}}{\text{(SD10\_SDT12/20au)}}  - \alpha_{\text{SM}}{\text{(SD12/20au)}} \\
 &+& \alpha_{\text{SM}}{\text{(SD10\_SDTQ12/5au)}}  - \alpha_{\text{SM}}{\text{(SD10\_SDT12/5au)}}
\end{eqnarray*}
The first of these two corrections -- the one due to combined Triple excitations -- amounts to $6.8$\% 
of the DCHF value. The second -- due to combined Quadruple excitations -- amounts to $2.4$\% of
the DCHF value. No higher excitations from the valence shells make a contribution larger than about 
$0.3$\%.

The uncertainty estimate for Hg results from adding individual uncertainties for the nuclear-density 
model ($3$\%), atomic basis set ($2$\%), higher excitation ranks ($4$\%), and the Breit interaction 
($1$\%).

Our final result is in agreement with the CI+MBPT result from reference \cite{dzuba_flambaum_PRA2009}.
These two results, however, are in disagreement with the coupled cluster value from 
Ref. \cite{sahoo_das-Hg_PRL2018}
even considering the estimated uncertainties. We have taken into account the leading higher excitations 
leading to a decrease of $\alpha_{\text{SM}}$ on the absolute, i.e., those which are also accounted
for in the coupled cluster expansion of Ref. \cite{sahoo_das-Hg_PRL2018}. We have in recent work
\cite{Fleig_Skripnikov2020} demonstrated that our present approach and strategy yields results of
similar quality as does a coupled cluster expansion of the wavefunction. Also, our total correlation
effect of around $17$\% for Hg is to be contrasted with around $39$\% correlation effect according
to Ref. \cite{sahoo_das-Hg_PRL2018}, the latter of which is unusually large for Schiff-moment interactions.

\begin{table}[h]
 \caption{\label{TAB_Hg:final-basis} Atomic Schiff moment interaction constant for Hg calculated at Hartree-Fock level
	including core contribution with Gaussian nuclear density \cite{Visscher_Dyall_nuclcha} 
	for studying electronic-basis-set convergence; $E_{\text{ext}}$ is set to $0.00024$ a.u.\\
        Augmented basis sets are built from Dyall's QZ set including diffuse and correlating functions.\\ 
           }

 \vspace*{0.3cm}
 \hspace*{-1.3cm}
\begin{tabular}{l|cc}
	Model & $\alpha_{\text{SM}}~ \left[ 10^{-17} \frac{e \text{cm}}{e \text{fm}^3} \right]$ & $\varepsilon_{\text{DCHF}}$ [a.u.]    \\ \hline
	DZ                                                     &$ 6.480$   & $-19648.85451859$  \\
	TZ                                                     &$-1.267$   & $-19648.89380174$  \\
	QZ-34s30p                                              &$-2.690$   & $-19648.88766826$  \\
	sp-densified QZ-67s59p                                 &$-2.898$   & $-19648.88651484$  \\\hline
	sp-densified+1sp QZ-69s61p                             &$-2.887$   & $-19648.88727782$  \\ \hline 
	sp-densified+2sp QZ-71s63p                             &$-2.887$   & $-19648.88727627$  \\
	sp-densified+3sp QZ-73s65p                             &$-2.884$   & $-19648.88727625$ \\ 
	sp-densified+4sp QZ-75s67p                             &$-2.896$   & $-19648.88727630$ \\ 
	sp-densified+5sp QZ-77s69p                             &$-2.897$   & $-19648.88727619$ \\ 
	sp-densified+6sp QZ-79s71p                             &$-2.900$   & $-19648.88727624$ \\ 
	sp-densified+7sp QZ-81s73p                             &$-2.886$   & $-19648.88727628$ \\ 
	sp-densified+8sp QZ-83s75p                             &$-2.886$   & $-19648.88727631$ \\ \hline
	Dzuba {\it{et al.}}\cite{Schiff_PRA_2002}              &$-2.8  $   &        -          \\
\end{tabular}
\end{table}

\clearpage

\begin{table}[h]
 \caption{\label{TAB:SCH_HG} Atomic Schiff moment interaction constant for Hg including electron correlation
	effects and the core contribution using the Dyall-cvQZ-69s61p basis set and a Gaussian nuclear density 
	\cite{Visscher_Dyall_nuclcha}; $E_{\text{ext}}$ is set to $0.00024$ a.u.
           }

 \vspace*{0.3cm}
 \hspace*{-1.3cm}
\begin{tabular}{l|c}
	Model              &  $\alpha_{\text{SM}}~ \left[ 10^{-17} \frac{e \text{cm}}{e \text{fm}^3} \right]$ \\ \hline
	DCHF                         &  $-2.887$   \\ \hline
	SD2/10au                     &  $-2.597$   \\
	SD2/20au                     &  $-2.599$   \\
	SD2/50au                     &  $-2.598$   \\ \hline
	SD12/10au                    &  $-2.614$   \\
	SD12/20au                    &  $-2.621$   \\
	SD12/50au                    &  $-2.623$   \\
	SDT12/10au                   &  $-2.426$   \\
	SD10\_SDT12/5au              &  $-2.408$   \\
	SD10\_SDT12/10au             &  $-2.420$   \\
	SD10\_SDT12/20au             &  $-2.425$   \\
	SD10\_SDTQ12/5au             &  $-2.339$   \\
	S6\_SD18/20au                &  $-2.599$   \\
	SD18/20au                    &  $-2.608$   \\ \hline
	SD20/10au                    &  $-2.568$   \\
	SD20/20au                    &  $-2.590$   \\ \hline
	S20\_SD32/20au               &  $-2.649$   \\ 
	SD32/20au                    &  $-2.696$   \\ \hline
	SD34/20au                    &  $-2.666$   \\ \hline
	{\bf{Final present}}         &  {\bf{-2.40$\pm$0.24}} \\ \hline
        Dzuba {\it{et al.}}\cite{dzuba_flambaum_PRA2009}       &$-2.6 $  \\
        Sahoo {\it{et al.}}\cite{sahoo_das-Hg_PRL2018}         &$-1.77 \pm 0.06$ 
  \end{tabular}
\end{table}

\subsection{{TlF}}

Both in Ref. \cite{Quiney_PTodd_PRA1997} and in our present work a careful and extensive 
optimization of atomic Gaussian basis sets for the Tl atom in TlF has been carried out. 
As to be seen in Table \ref{TAB_TlF:final-basis} the DCHF result for our final $sp$-densified+$1sp$ 
QZ-69s63p basis set differs from the corrected literature result by Quiney et al.
\cite{Flambaum-Dzuba_TranPRA2020,Quiney_PTodd_PRA1997} by only $2.2$\%. We consider this 
agreement as a further confirmation of the reliability of our technique of basis-set
optimization. The DCHF result by Skripnikov et al. \cite{Skripnikov_actinideSchiff_2020} is 
within about $4$\% of the result by Quiney et al.

\begin{table}[h]
 \caption{\label{TAB_TlF:final-basis} Molecular Schiff moment interaction constant for TlF calculated at Hartree-Fock level
including core contribution with Gaussian nuclear density \cite{Visscher_Dyall_nuclcha} at $R = 3.94$ a.u. \cite{Huber:1979} 
for studying electronic-basis-set convergence.
Augmented basis sets are built from Dyall's QZ set including diffuse and correlating functions.\\ 
           }

 \vspace*{0.3cm}
 \hspace*{-1.3cm}
\begin{tabular}{l|ccc}
	Model & $ W_{\text{SM}}({\text{Tl}})$ [a.u.] & $\varepsilon_{\text{DCHF}}$  [a.u.]   \\ \hline
	QZ-34s31p                       & 42877 &   $-20374.47704191$  \\
	sp-densified QZ-67s61p          & 30737 &   $-20374.47575145$  \\ \hline
	{\bf{sp-densified+1sp QZ-69s63p}}      & {\bf{45419}} &   $-20374.47660904$  \\\hline
	sp-densified+2sp QZ-71s65p      & 45540 &   $-20374.47660743$  \\ 
	sp-densified+3sp QZ-73s67p      & 45584 &   $-20374.47660800$ \\ 
	sp-densified+4sp QZ-75s69p      & 45602 &   $-20374.47660786$ \\  
	sp-densified+5sp QZ-77s71p      & 45578 &   $-20374.47660735$ \\ 
	sp-densified+6sp QZ-79s73p      & 45594 &   $-20374.47660803$ \\ 
	sp-densified+7sp QZ-81s75p      & 45602 &   $-20374.47660837$ \\ 
	sp-densified+8sp QZ-83s77p      & 45584 &   $-20374.47660780$ \\ \hline
	Quiney et al. (DCHF)\footnote{Value from Ref. \cite{Quiney_PTodd_PRA1997} corrected in
	Ref. \cite{Flambaum-Dzuba_TranPRA2020} for the use of a more accurate operator for the
	Schiff-moment interaction}      & 46444 &                     \\
	Skripnikov et al. (DCHF) \cite{Skripnikov_actinideSchiff_2020}  & 48377 &             \\ \hline
\end{tabular}
\end{table}

\begin{table}[h]
 \caption{\label{TAB:SCH_TLF} Molecular Schiff moment interaction constant for the thallium nucleus in
	TlF including electron correlation effects and the core contribution using the 
	Dyall-cvQZ-69s63p basis set (denoted cvQZ+ in the Table) 
	and a Gaussian nuclear density \cite{Visscher_Dyall_nuclcha} at $R = 3.94$ a.u. \cite{nist_tlf}}
\begin{tabular}{l|ccl}
 Basis set/Model        &  $W_{\text{SM}}({\text{Tl}})$~ [a.u.] \\ \hline
 cvQZ/DCHF                                           &     $42877$     \\
 cvQZ+/DCHF                                          &     $45419$     \\ \hline
 cvQZ+/SD8\_5.5au                                    &     $40779$     \\ 
 cvQZ+/SD8\_10au                                     &     $41198$     \\ 
 cvQZ+/SD8\_20au                                     &     $41431$     \\ 
 cvQZ+/SD8\_40au                                     &     $41438$     \\ 
 cvQZ+/SDT8\_5.5au                                   &     $39954$     \\
 cvQZ+/SDT8\_10au                                    &     $40314$     \\ 
 cvQZ+/SDT8\_20au                                    &     $40495$     \\ 
 cvQZ+/0.8auSDTQ8\_SDT8\_5.5au                       &     $38863$      \\ \hline
 cvQZ+/SD10\_10au                                    &     $41584$     \\
 cvQZ+/SDT10\_10au                                   &     $40544$     \\
 cvQZ+/SD16\_10au                                    &     $41762$     \\ \hline
 cvQZ+/SD18\_10au                                    &     $41838$     \\ \hline
 cvQZ+/SD20\_10au                                    &     $41858$     \\ \hline
{\bf{Final present}}                                 &  {\bf{39967$\pm$3600}} \\ \hline
Flambaum {\it{et al.}}\cite{Flambaum-Dzuba_TranPRA2020,Petrov_TlF_PRL2002} (CC)  &     $40539$     \\
Skripnikov et al. (CCSD(T)) \cite{Skripnikov_actinideSchiff_2020}  & $37192$    \\ 
Abe      {\it{et al.}}\cite{Abe_Schiff_2020}            (CC)  &     $41136$     \\
\end{tabular}
\end{table}

In Table \ref{TAB:SCH_TLF} we compile correlated Schiff-moment interactions for TlF using various
CI models. The main correction to the DCHF result comes from valence correlations among the $6s$ (Tl)
and $2p$ (F) electrons (model SD8), quenching the interaction constant by nearly $9$\%. Full
Triple excitations out of these shells further reduce the value by almost $2$\%
and in addition by a similar amount when a leading set of Quadruple excitations is added
to the expansion, model 0.8auSDTQ8\_SDT8 (where up to four particles are allowed in shells below
an energy cutoff of $0.8$ a.u.).
Correlation contributions from the shells $1s,2s$ (F) and $5s,5p$ (Tl) are seen to be
small, amounting to an increase of $W_{\text{SM}}$ by only about $1.5$\%.

For TlF we obtain our final best result from a base value obtained with the model cvQZ+/SDT10\_10au
and adding corrections for Quadruple excitations from the valence shells, correlation
contributions from $1s$ (F) and $5s,5p$ (Tl) shells and a cutoff correction for the valence shell
correlations, according to
\begin{eqnarray*}
W_{\text{SM}}{\text{(Final)}} &=& W_{\text{SM}}({\text{cvQZ+/SDT10\_10au)}} \\
 &+& W_{\text{SM}}{\text{(cvQZ+/0.8auSDTQ8\_SDT8\_5.5au)}}  - W_{\text{SM}}{\text{(cvQZ+/SDT8\_5.5au)}} \\
 &+& W_{\text{SM}}{\text{(cvQZ+/SD20\_10au)}}  - W_{\text{SM}}{\text{(cvQZ+/SD10\_10au)}} \\
 &+& W_{\text{SM}}{\text{(cvQZ+/SD8\_40au)}}  - W_{\text{SM}}{\text{(cvQZ+/SD8\_10au)}}
\end{eqnarray*}
The uncertainty estimate for TlF results from adding individual uncertainties for the nuclear-density
model ($3$\%), atomic basis set ($2$\%), higher excitation ranks ($2$\%), inner-shell correlations
($1$\%) and the Breit interaction ($1$\%).

Our final best result is in agreement with both the operator-shifted results of Flambaum {\it{et al.}}
in Refs. \cite{Flambaum-Dzuba_TranPRA2020,Petrov_TlF_PRL2002} and the recent CCSD(T) calculation by
Skripnikov et al. \cite{Skripnikov_actinideSchiff_2020}.
\section{Conclusions}
\label{SEC:CONCL}
Using the ${\cal{P,T}}$-violating energy shift $\Delta\varepsilon$ from the most recent measurements on the present 
systems and our calculated interaction constants we can determine the nuclear Schiff moment $S$ itself, in the context 
of a single-source assumption. It results from the relation
\begin{equation}
 \Delta\varepsilon = 2 W_{\text{SM}}\, S
\end{equation}
where using our final central value for $W_{\text{SM}}$ from Table \ref{TAB:SCH_TLF} and the measured frequency
shift of $(1.4 \pm 2.4) \times 10^{-4}$ Hz from Ref. \cite{ChoSangsterHinds_TlF_PRA1991} yields
\begin{equation}
	S(^{205}{\text{Tl}}) = (3.9 \pm 6.8) \times 10^{-11}\, e\, {\text{fm}}^3
	\label{EQ:S_TL}
\end{equation}
for the Schiff moment of the $^{205}{\text{Tl}}$ nucleus. The CENTREX collaboration \cite{CENTREX_2021} aims at a
significant increase in sensitivity to hadronic ${\cal{T}}$-violating fundamental interactions which -- combined
with the recent results for $W_{\text{SM}}$ -- will lead to stronger constraints on the nuclear Schiff moment in
case of a null measurement with tighter uncertainties.

The limit on the nuclear Schiff moment can be used to infer limits on the CPV pion-nucleon couplings, the QCD
$\Theta$ parameter and chromo-EDMs following the relations in Ref. \cite{Flambaum-Dzuba_TranPRA2020}.

A stronger constraint than the one in Eq. \ref{EQ:S_TL} can be placed on the Schiff moment of the 
{$^{199}$Hg} nucleus. Using the upper bound on the Hg EDM of
\begin{equation}
	\left|d_{\text{Hg}}\right| < 7.4 \times 10^{-30}\, e\, {\text{cm}}
\end{equation}
from Ref. \cite{Heckel_Hg_PRL2016} and our central value for the Schiff-moment interaction from Table
\ref{TAB:SCH_HG} yields an upper bound to the Hg Schiff moment
\begin{equation}
	\left|S_{\text{Hg}}\right| < 3.1 \times 10^{-13}\, e\, {\text{fm}}^3
\end{equation}
This is the same value as the one proposed in Ref. \cite{Heckel_Hg_PRL2016} where an average over 
various uncorrelated and correlated Schiff-moment interactions from the literature had been used.
In the present case a rigorously calculated interaction parameter $\alpha_{\text{SM}}$(Hg) including 
the effects of 
interelectron correlations in the Hg atom ground state replaces that average value which it 
happens to match.
The nuclear Schiff moment of {$^{129}$Xe} and {$^{199}$Hg} has recently been calculated as a function
of the strong $\pi$-meson-nucleon interaction constants \cite{Yanase_Shimizu_Schiff2020}. Combined
with these dependencies our results can be used to constrain these interaction constants.

\section{Acknowledgments}
\label{SEC:ACK}
We thank Vladimir Dzuba (Sydney) for numerous helpful discussions and for sharing details 
of his calculations with us.
\bibliographystyle{unsrt}

\clearpage
\begin{appendix}
\section{Details on Basis-Set Optimization for Xe}

\begin{table}[h]
	\caption{\label{TAB_Xe:nsp-densified} Atomic Schiff moment interaction constant $\alpha_\text{SM}$ for Xe calculated at Hartree-Fock level
	including core contribution with Gaussian nuclear density \cite{Visscher_Dyall_nuclcha} 
	for studying electronic-basis-set convergence using several $s$ and $p$ spaces densifications 
	and addition of denser and more diffuse $(s,p)$ function pairs; $E_{\text{ext}}$ is set to $0.0003$ a.u.
        Augmented basis sets are built from Dyall's QZ with diffuse and correlating functions.
	+1sp  means one denser $s$ and $p$ function, one more diffuse $s$ and $p$ function have been added to the basis set.\\
           }

 \vspace*{0.3cm}
 \hspace*{-1.3cm}
\begin{tabular}{l|c|c|l}
	Model & $(s,p)$ space&{$\alpha_{\text{SM}}~ \left[ 10^{-17} \frac{e \text{cm}}{e \text{fm}^3} \right]$}& $\varepsilon_{\text{DCHF}}$ [a.u.]    \\ \hline
	QZ                          &34s28p  &0.318&$-7446.895409376$  \\
	QZ +1sp*                    &36s30p  &0.369&$-7446.895386026$  \\
	Simple sp-densified QZ      &67s55p  &0.314&$-7446.895379750$  \\
	Simple sp-densified QZ +1sp &69s57p  &0.373&$-7446.895401869$  \\
	Double sp-densified QZ      &133s109p&0.362&$-7446.895392349$  \\ 
	Double sp-densified QZ +1sp &135s111p&0.359&$-7446.895401000$  \\ 
	Double sp-densified QZ +2sp &137s113p&0.354&$-7446.895401764$  \\ 
	Double sp-densified QZ +3sp &139s115p&0.369&$-7446.895401848$  \\ 
	Double sp-densified QZ +4sp &141s117p&0.372&$-7446.895401815$  \\ 
	Double sp-densified QZ +5sp &143s119p&0.375&$-7446.895401836$  \\ 
	Double sp-densified QZ +6sp &145s121p&0.376&$-7446.895401842$  \\ 
	Double sp-densified QZ +7sp &147s123p&0.375&$-7446.895401832$  \\ 
	Double sp-densified QZ +8sp &149s125p&0.376&$-7446.895401826$  \\ 
	Double sp-densified QZ +9sp &151s127p&0.376&$-7446.895401837$  \\ 
	Double sp-densified QZ +10sp&153s129p&0.376&$-7446.895401813$  \\ 
	Triple sp-densified QZ      &265s217p&0.051&$-7446.895394063$  \\ 
	Triple sp-densified QZ +1sp &267s219p&0.684&$-7446.895399110$  \\ \hline
\end{tabular}
\end{table}
{\tiny
}
\begin{table}[h]
 \caption{\label{TAB_Xe:dense_VS_dense+diffuse} Atomic Schiff moment interaction constant $\alpha_\text{SM}$~$\left[ 10^{-17} \frac{e \text{cm}}{e \text{fm}^3} \right]$ for Xe calculated at Hartree-Fock level
	including core contribution with Gaussian nuclear density \cite{Visscher_Dyall_nuclcha} 
	for studying electronic-basis-set convergence by addition of denser $(s,p)$ pairs 
	versus addition of denser $(s,p)$ and more diffuse $(s,p)$ pairs; $E_{\text{ext}}$ is set to $0.0003$ a.u.
        Augmented basis sets are built from Dyall's QZ set with diffuse and correlating functions.\\ 
           }
 \vspace*{0.3cm}
 \hspace*{-1.3cm}
\begin{tabular}{l||c|c|c||c|c|c}
	&\multicolumn{3}{c||}{Addition of $(s,p)$ dense pairs} & \multicolumn{3}{c}{Addition of $(s,p)$ dense/diffuse pairs} \\ \hline
	Model &$(s,p)$ space& {$\alpha_{\text{SM}}$}& $\varepsilon_{\text{DCHF}}$ [a.u.] &$(s,p)$ space&{$\alpha_{\text{SM}}$}& $ \varepsilon_{\text{DCHF}}$  [a.u.]   \\ \hline
	QZ                       &34s28p  &0.318& $-7446.895409376$   &34s28p  &0.318&$-7446.895409376$  \\
	QZ +1sp                  &35s29p  &0.368& $-7446.895385894$   &36s30p  &0.369&$-7446.895386026$  \\
	QZ +2sp                  &36s30p  &0.369& $-7446.895384072$   &38s32p  &0.369&$-7446.895384244$  \\
	QZ +3sp                  &37s31p  &0.369& $-7446.895383784$   &40s34p  &0.370&$-7446.895383956$  \\
	QZ +4sp                  &38s32p  &0.369& $-7446.895383738$   &42s36p  &0.371&$-7446.895383915$  \\
	QZ +5sp                  &39s33p  &0.371& $-7446.895383742$   &44s38p  &0.370&$-7446.895383987$  \\
	QZ +6sp                  &40s34p  &0.370& $-7446.895383941$   &46s40p  &0.370&$-7446.895384141$  \\
	QZ +7sp                  &41s35p  &0.370& $-7446.895383410$   &48s42p  &0.372&$-7446.895384163$  \\
	QZ +8sp                  &42s36p  &0.370& $-7446.895383571$   &50s44p  &0.376&$-7446.895382476$  \\ \hline
	sp-densified QZ          &67s55p  &0.314& $-7446.895379750$   &67s55p  &0.314&$-7446.895379750$  \\          
	sp-densified QZ+1sp      &68s56p  &0.378& $-7446.895401868$   &69s57p  &0.373&$-7446.895401869$  \\          
	sp-densified QZ+2sp      &69s57p  &0.375& $-7446.895401802$   &71s59p  &0.375&$-7446.895401810$  \\          
	sp-densified QZ+3sp      &70s58p  &0.376& $-7446.895401801$   &73s61p  &0.375&$-7446.895401762$  \\
	sp-densified QZ+4sp      &71s59p  &0.376& $-7446.895401801$   &75s63p  &0.375&$-7446.895401779$  \\
	sp-densified QZ+5sp      &72s60p  &0.378& $-7446.895401802$   &77s65p  &0.376&$-7446.895401790$  \\
	sp-densified QZ+6sp      &73s61p  &0.375& $-7446.895401802$   &79s67p  &0.375&$-7446.895401745$  \\
	sp-densified QZ+7sp      &74s62p  &0.375& $-7446.895401799$   &81s69p  &0.374&$-7446.895401770$  \\
	sp-densified QZ+8sp      &75s63p  &0.376& $-7446.895401800$   &83s71p  &0.375&$-7446.895401790$  \\ \hline
\end{tabular}
\end{table}

\end{appendix}

\end{document}